\documentclass[aps,pra,twocolumn,showpacs,floatfix]{revtex4}

\usepackage{amsmath,float,graphicx,subfigure}

\begin{document}

\title{Effects of trap anisotropy on impurity scattering regime in a
Fermi gas}

\author{P. Capuzzi} 
\email{capuzzi@sns.it}
\author{P. Vignolo} 
\email{vignolo@sns.it}
\author{M. P. Tosi}
\email{tosim@sns.it}
\affiliation{NEST-INFM and
Classe di Scienze, Scuola Normale Superiore, I-56126 Pisa, Italy}

\begin{abstract}
We evaluate the low-lying oscillation modes and the ballistic
expansion properties of a harmonically trapped gas of fermionic
$^{40}$K atoms containing thermal $^{87}$Rb impurities as functions of
the anisotropy of the trap.  Numerical results are obtained by solving
the Vlasov-Landau equations for the one-body phase-space distribution
functions and are used to test simple scaling Ansatzes.  Starting from
the gas in a weak impurity-scattering regime inside a spherical trap,
the time scales associated to motions in the axial and azimuthal
directions enter into competition as the trap is deformed to an
elongated cigar-like shape. This competition gives rise to coexistence
of collisionless and hydrodynamic behaviors in the low-lying surface
modes of the gas as well as to a dependence of the aspect ratio of the
expanding cloud on the collision time.

\end{abstract}

\maketitle  

\section{\label{sec:intro}Introduction}
Experiments on ballistic expansion and collective-mode excitations
have provided important diagnostic tools in the study of
quantum-degenerate atomic and molecular gases (for a recent review see
~\cite{Minguzzi2004a}).  A milestone example is the interpretation of
the anisotropic expansion of a cloud of ultracold bosons as the
smoking gun of Bose-Einstein condensation in a trapped dilute gas
\cite{Anderson1995a, Bradley1995a, Davis1995b}. The later achievement
of degenerate Fermi gases \cite{DeMarco1999a} and boson-fermion
mixtures \cite{Truscott2001a,Schreck2001a,Goldwin2002a} has opened up
the possibility to experimentally observe rich quantum phase diagrams
and to study the interplay between external confinement and
interspecies collisions.  More recently, the expansion behavior of
strongly interacting fermion clouds has been the object of experiments
aimed at probing novel superfluid states on the crossover from a BCS
superfluid to a condensate of molecular dimers
\cite{Ohara2002b,Greiner2003a,Chin2004a}.

Collective modes provide a direct measure of the collisionality of a
dilute quantum gas.  Monopolar and quadrupolar modes have been
proposed as markers of the approach of quantum phase transitions
in binary mixtures of Bose-Einstein condensed gases (BEC's)
\cite{Graham1998a}, in boson-fermion mixtures \cite{capuzzi2003a,
capuzzi2003b}, and in fermion mixtures across the BCS-BEC 
crossover \cite{Kinast2004a,Bartenstein2004a,Kinast2004b}. 
The dynamical transition from collisionless to
hydrodynamic behavior in fermion mixtures has been followed
experimentally \cite{Gensemer2001a} and numerically \cite{Toschi2003a}
in two-component fermion mixtures by studying their dipolar
oscillation modes. The role played in this context by mobile
impurities inside a fermion gas under spherical confinement has also
been studied by numerical means \cite{Capuzzi2004b}.

In the present work we examine how the anisotropy of the trap affects
the low-lying oscillation modes and the ballistic expansion of a gas
of fermionic $^{40}$K atoms containing a small concentration of
thermal $^{87}$Rb atoms.  We numerically solve the Vlasov-Landau
equations for the evolution of the phase-space distribution functions
within a particle-in-cell approach and compare the results with simple
scaling Ansatzes. Our results demonstrate that in a cigar-shaped
harmonic confinement, where two different confinement frequencies
determine two different time scales, collisionless and hydrodynamic
behaviors can coexist in the low-lying collective excitations of the
gas and that the aspect ratio of the expanding cloud shows a
non-monotonic dependence on the anisotropy of the trap.

The paper is organized as follows. In Sect.\ \ref{sec:mix} we
introduce the mixture under study and the basic equations that will be
used to describe its dynamics. Section \ref{sec:osc_modes} analyzes
the monopolar, dipolar, and quadrupolar oscillations of the gas as
functions of the anisotropy of the confining potential, while Sect.\
\ref{sec:expan} gives a discussion of the dynamics of the free
expansion of the fermion cloud. Finally, Sect.\ \ref{sec:summary}
presents a summary and the main conclusions of our work.

\section{\label{sec:mix}The mixture}
We consider a gas of fermionic atoms of mass $m_F$ confined inside
an axially symmetric harmonic trap of the form
\begin{equation}
V_{F}(\mathbf{r}) =
\frac{1}{2}\,m_F\,\omega_{F,\perp}^2 
\left(x^2 + y^2 + \lambda^2\,z^2\right), 
\label{eq:VF}
\end{equation}
where $\omega_{F,\perp}$ is the angular trap frequency for motions
along the $\hat{x}$ and $\hat{y}$ directions and $\lambda =
\omega_{F,\,z}/\omega_{F,\,\perp}$ is the anisotropy parameter.
Potentials with $\lambda \ll 1$ generate cigar-shaped density
profiles. The fermions are wholly spin-polarized and hence at very low
temperature the Pauli principle quenches collisions among them. A
second component must then be added in order to thermalize the gas and
to drive its collisionality.

We dope the fermion gas with
a small number of bosons having larger atomic mass $m_B$. 
We simply think
of them as impurities confined by the harmonic potential
\begin{equation}
V_{B}(\mathbf{r}) =
\frac{1}{2}\,m_B\,\omega_{B,\perp}^2
\left(x^2 + y^2 + \lambda^2\,z^2\right),
\label{eq:VB}
\end{equation}
where $\omega_{B,\perp}$ is the angular trap frequency in the
azimuthal plane and we have assumed that bosons and fermions share the
same trap anisotropy $\lambda$. The difference in atomic masses and
particle numbers causes a much lower quantum degeneracy temperature
for the bosons and we can therefore consider them as an uncondensed cloud, even
in the presence of highly degenerate fermions.

The dynamics of the mixture is described through the one-body
distribution functions $f^{(F, B)}(\mathbf{r}, \mathbf{p}, t)$ in the
Boltzmann approximation. Their evolution is governed by the
Vlasov-Landau kinetic equations (VLE),
\begin{equation}
\partial_t f^{(j)}+\dfrac{\bf p}{m_j}\cdot{\bf \nabla}_{{\bf r}}
f^{(j)}- {\bf \nabla}_{{\bf r}} U^{(j)}\cdot {\bf \nabla}_{{\bf
p}}f^{(j)}= C[f^{(F)},f^{(B)}]
\label{vlasov}
\end{equation}
where the Hartree-Fock effective potential is $U^{(j)}({\bf r},t)=
V_{j} ({\bf r})+g n^{(\overline{j})}({\bf r},t)$ with $\overline{j}$
denoting the species different from $j$.  Here we have set $g=2\pi
\hbar^2 a/m_r$ with $a$ being the $s$-wave scattering length of a
fermion-boson pair and $m_r$ its reduced mass, and $n^{(j)}({\bf
r},t)$ is the spatial density given by integration of $f^{(j)}({\bf
r},{\bf p},t)$ over the momentum degrees of freedom. Since we deal
with low concentrations of impurities we have neglected
impurity-impurity interactions. In addition, collisions between
spin-polarized fermions are negligible at low temperature and thus the
collision integral $C$ in Eq.\ (\ref{vlasov}) involves only collisions
between fermions and impurities. This is given by
\begin{eqnarray}
C &=& \frac{\sigma}{4\pi(2\pi\hbar)^3}\int\!\! d^3p_2\,d\Omega_f
v[(1-f^{(F)})(1+f_2^{(B)})f_3^{(F)} f_4^{(B)} \nonumber \\ &&  -
f^{(F)}f^{(B)}_2(1-f_3^{(F)}) (1+f_4^{(B)})],
\label{integralcoll}
\end{eqnarray}
where $f^{(j)}\equiv f^{(j)}({\bf r},{\bf p},t)$ and $f_i^{(j)}\equiv
f^{(j)}({\bf r},{\bf p}_i,t)$, $d\Omega_f$ is the element of solid
angle for the outgoing relative momentum $\mathbf{p}_3-\mathbf{p}_4$,
$v=|\mathbf{v}-\mathbf{v}_2|$ is the relative velocity of the incoming
particles, and $\sigma = 4\pi a^2$ is the scattering
cross-section. The collision satisfies conservation of momentum
($\mathbf{p}+\mathbf{p}_2 = \mathbf{p}_3 + \mathbf{p}_4$) and energy
($\varepsilon + \varepsilon_2 = \varepsilon_3 + \varepsilon_4$), with
$\varepsilon_j= p_j^2/2m_j + U^{(j)}$.

The solution of Eqs.\ (\ref{vlasov}) is carried out by using a
numerical approach based on particle-in-cell plus Monte Carlo sampling
techniques, which allows us to evaluate the kinetics of such systems
down to $T\simeq 0.1\,T_F$. The technical details of the method have
been given elsewhere \cite{Toschi2003a,Capuzzi2004b,Succi2003a}.

In the following we shall focus on the dependence of the dynamics of a
mixture of $N_F=10^4$ $^{40}$K atoms and $N_B=10^2$ $^{87}$Rb atoms on
the anisotropy $\lambda$ of the traps at fixed values of the average
trap frequencies $\bar\omega_F=(\omega_{F,\perp}^2\omega_{F,z})^{1/3}$
and $\bar\omega_B=(\omega_{B,\perp}^2\omega_{B,z})^{1/3}$. These are
taken as the geometric averages of the trap frequencies in the
experiments carried out at LENS on $^{40}$K-$^{87}$Rb mixtures
\cite{Ferlaino2003a}, {\it i.e.} we set $\bar{\omega}_F=2\pi\times
134.1$ s$^{-1}$ and $\bar{\omega}_B=2\pi\times 91.2$ s$^{-1}$.  We fix
the temperature at $T=0.2\,T_F$, with $T_F = \hbar
\bar{\omega}_F\,(6N_F)^{1/3}/k_B$ being the Fermi degeneracy
temperature for noninteracting fermions. This also corresponds to
$T\simeq 2.7\,T_{\text{BEC}}$ where $T_{\text{BEC}}=
0.94\hbar\bar{\omega}_B\,(N_B)^{1/3}/k_B$ is the condensation
temperature for the noninteracting Bose component. Finally, we assume
a repulsive $s$-wave scattering length $a=2000$ Bohr
radii. 
This choice of the fermion-boson scattering length is dictated by 
computational convenience, but still leaves the mixture at $\lambda=1$ in
a collisionless-to-intermediate scattering regime as for the
$^{40}$K-$^{87}$Rb mixtures studied at LENS.

\section{\label{sec:osc_modes}Oscillation modes}
\subsection{Dipolar oscillations}
The collisional state of the gas is revealed by the behavior of the
frequencies and damping rates of collective excitations as functions
of the collision frequency. Briefly, for dipolar modes the
hydrodynamic behavior is signalled by a common frequency of
oscillation of the two species and by a decrease of the damping rate
with increasing collision frequency. Conversely the collisionless
regime, which is attained at low collisionality, is characterized by
different oscillation frequencies and by growing damping rates. The
collision rate can be evaluated either from a numerical simulation
which actually counts the number of collisions at each time step, or
by direct integration of the collision integral over momenta. Pauli
blocking strongly quenches collisions at the temperatures of present
interest and its handling in numerical studies requires suitably
adapted methods of Monte Carlo sampling \cite{Toschi2003a,
Capuzzi2004b,Succi2003a}.

\begin{figure}
\includegraphics[width=\columnwidth,clip=true]{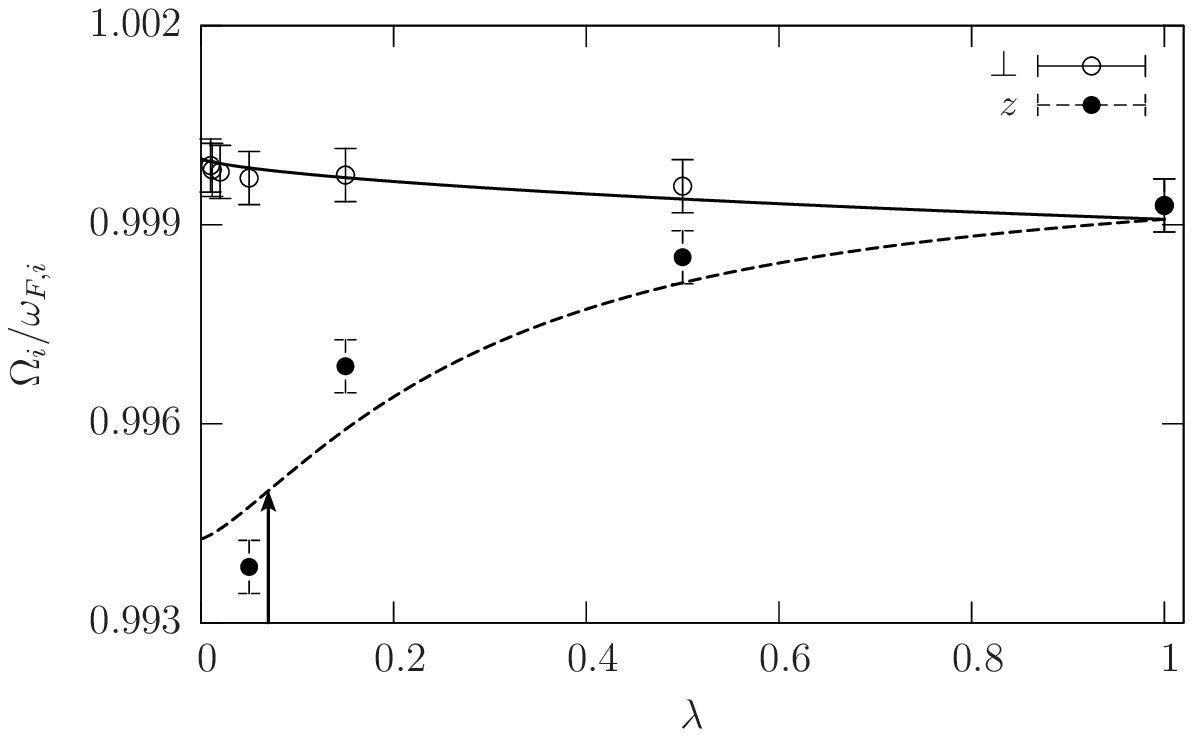}\\[3pt]
\includegraphics[width=\columnwidth,clip=true]{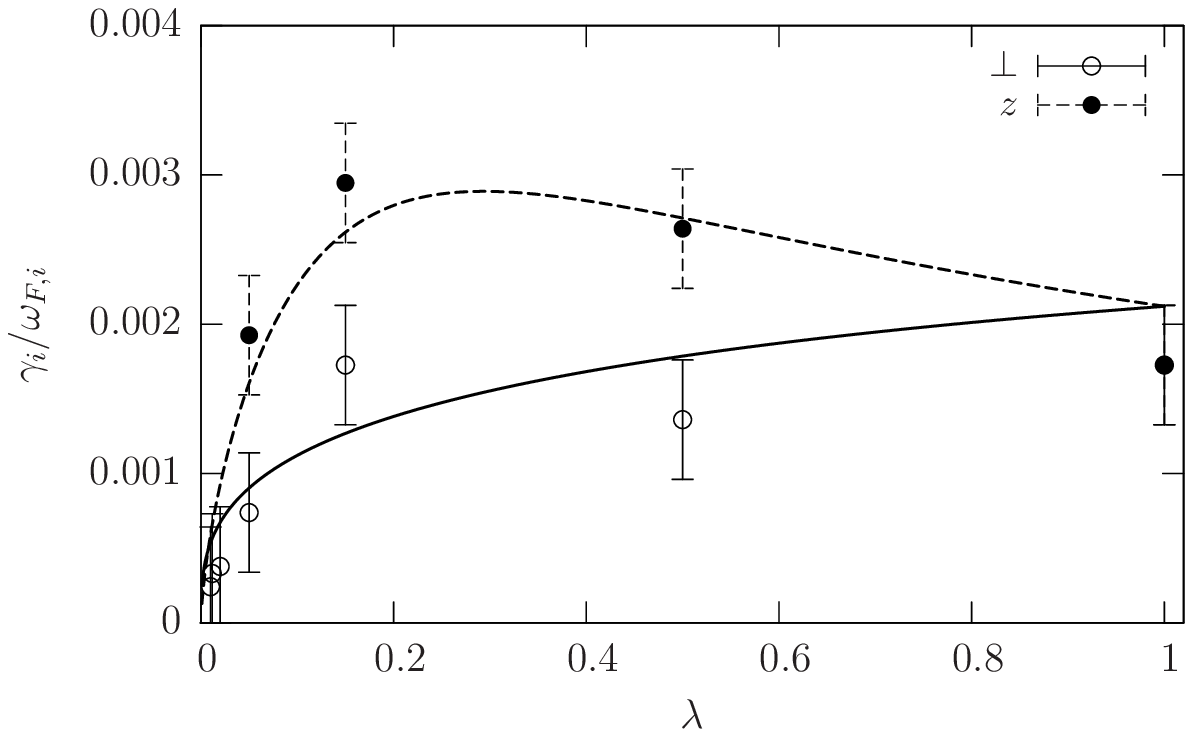}
\caption{\label{fig:dipo_comp}Fermion dipolar frequencies $\Omega_{i}$
(top panel) and damping rates $\gamma_{i}$ (bottom panel) in units of
$\omega_{F,i}$, as functions of the trap anisotropy $\lambda$.  The
results of the simulation (symbols with error bars) are compared with
those obtained from Eqs.\ (\ref{eq:dyndip_MF}): open circles and solid
lines denote radial oscillations, while full circles and dashed lines
refer to motions along the $\hat{z}$ axis.  The arrow in the top panel
marks the locking point given in Eq.\ (\ref{eq:lock}).}
\end{figure}

In the numerical simulation we excite dipolar modes by initially
shifting the fermionic density profile by $a_{ho}=\sqrt{\hbar/
(m_F\omega_{F,\perp})}$ in either the axial or radial direction ($\hat
z$ and $\hat x$, say) and then fit the time evolution of the fermionic
center-of-mass coordinates with the functions $\cos(\Omega_i\,t+\phi)
\exp(-\gamma_i\,t)$. From these fits we extract the oscillation
frequencies $\Omega_i$ and the damping rates $\gamma_i$ along the two
directions.  The results obtained from the numerical solution of the
Vlasov-Landau equations are shown in Fig.\ \ref{fig:dipo_comp}.  The
error bars in the plot have been estimated from the standard
deviations of $\Omega_i$ and $\gamma_i$ in the fitting process.

Starting from $\lambda=1$ and increasing the anisotropy towards an
elongated cigar-shaped trap, the frequencies of the dipolar
oscillations show different behaviors along the axial and the radial
direction. The value of $\Omega_{\perp}/\omega_{F,\perp}$ increases
slightly towards unity as $\lambda$ is decreased, whereas
$\Omega_z/\omega_{F,z}$ decreases with $\lambda$ towards an
appreciably lower value. At the same time both damping rates tend to
vanish, although that for axial oscillations appears to go through
a broad maximum before doing so. These behaviors indicate that in
strongly elongated traps the dynamics of the Fermi gas is
collisionless in radial dipolar oscillations, but collisional in axial
ones.

The coexistence of collisionless and hydrodynamic behaviors in the
dipolar oscillations as presented above is supported by the solution
of scaling equations in the classical limit. Within a classical model
one can write a set of equations for the coupled motions of the
centers of mass of fermions and bosons in the small-oscillation regime
\cite{Gensemer2001a,Ferlaino2003a}.  If we include mean-field effects,
the equations of motion in the $i$-direction for the center-of-mass
coordinates and for the relative coordinates are
\begin{equation}
\left\{
\begin{aligned}
\ddot{x}_{{\rm CM},i} &= -\omega_{\text{hd},i}^2\,x_{{\rm CM},i}
-M_r\Delta\omega^2\,x_{r,i}/M  \\
\ddot{x}_{r,i} &= -\Delta\omega^2\,x_{{\rm CM},i}
-\omega^2_{r,i}\,x_{r,i}+\omega^2_{{\rm mf},i}\,x_{r,i}
-\omega_Q\,\dot{x}_{r,i}
\end{aligned}
\label{eq:dyndip_MF}
\right.
\end{equation}
where $x_{\text{CM},i}= ( m_F N_F x_{F,i} + m_B N_B x_{B,i})/M$ and
$x_{r,i} = x_{F,i} - x_{B,i}$.  We have defined
$\Delta\omega^2=\omega_{F,i}^2-\omega_{B,i}^2$ and $\omega^2_{r,i}=(
m_B N_B \omega_{F,i}^2 + m_F N_F \omega_{B,i}^2)/M$ with $M =  m_F N_F
+ m_B N_B $ and $M_r=m_B m_F N_B N_F/M$.  The
hydrodynamic frequencies are given by
\begin{equation}
\omega_{\text{hd},i} = \left(\frac{m_F\,N_F\,\omega_{F,i}^2 +
m_B\,N_B\,\omega_{B,i}^2}{M}\right)^{1/2} 
\label{eq:whd}
\end{equation}
and the collisional frequency
is $\omega_Q=4Q(m_B/N_F+m_F/N_B)/(3m_F+3m_B)$, $Q$ being the total 
number of collisions per
unit time defined as
\begin{equation}
Q = \frac{\sigma}{\pi^2}\,\dfrac{N_F\,N_B}{k_BT}\,
\left(\frac{k_x\,k_y\,k_z}{m_r}\right)^{1/2}\,.
\label{eq:Qdef}
\end{equation}
In Eq.\ (\ref{eq:Qdef})
$k_i=k_{F,\,i}\,k_{B,\,i}/(k_{F,\,i}+k_{B,\,i})$ are the effective
oscillator constants, with $k_{j,\,i}=m_j\,\omega_{j,\,i}^2$.  In the
large-$Q$ limit Eqs.\ (\ref{eq:dyndip_MF}) predict that the relative
motion of the two clouds is overdamped, so that fermions and bosons
oscillate together at the hydrodynamic frequencies.  For the system
parameters that we are using these are $\omega_{\text{hd},i} =
0.9942\,\omega_{F,i}$. While due to the low number of impurities the
value of $\omega_{\text{hd},i}$ is very close to the bare trap
frequency $\omega_{F,i}$, their difference can be amplified by, e.g.,
increasing the mass of the impurities or varying the trapping
frequency.

The mean-field correction to the bare frequency of the relative motion
in Eq.\ (\ref{eq:dyndip_MF}) is
\begin{equation}
\omega^2_{{\rm mf},i}=\dfrac{g}{M_r}\int d^3r\, \frac{\partial
n^{\scriptscriptstyle(F)}_0}{\partial x_i}\,\frac{\partial
n^{\scriptscriptstyle(B)}_0}{\partial x_i}, 
\label{eq:mfcorrection}
\end{equation}
where $n^{\scriptstyle(F,B)}_0(\mathbf{r})$ are the equilibrium
density profiles.  Equation (\ref{eq:mfcorrection}) extends the result
of Ref.\ \cite{Vichi1999a} to a general gaseous mixture.  According to
Eqs.\ (\ref{eq:dyndip_MF}) the mean-field correction is more important
in the collisionless limit and does not affect the value of the
hydrodynamic frequency. However, it shifts the value of
$Q^{\text{lock}}$ at which the two clouds become glued together. The
locking point can be estimated by looking for an overdamped
oscillation in the solution for the relative motion in Eqs.\
(\ref{eq:dyndip_MF}). If we neglect the coupling between the
center-of-mass and the relative motions we find
\begin{equation}
\sqrt{\omega^2_{r,i}-\omega^2_{{\rm
mf},i}} \simeq\left.\frac{\omega_Q}{2}\,\right\vert_{Q^{\text{lock}}}.
\label{eq:lock}
\end{equation}
We have estimated a value of
$\omega^2_{\text{mf},i}/\omega^2_{F,i}\sim 8\times 10^{-3}$ for the
isotropic trap and verified that this ratio is slightly decreasing
with $\lambda$.  Therefore the mean-field correction is negligible in
the present case and the locking point is fixed by the bare frequency
of the relative coordinate, namely
$\left.\omega_Q\right\vert_{Q^{\text{lock}}}\simeq 2\omega_{r,i}
\simeq 1.4\,\omega_{F, i}$.

The classical model does not contain the effects of Pauli blocking at
low temperature, so that we have taken $Q$ as a single fitting
parameter at all values of $\lambda$. The results obtained from Eqs.\
(\ref{eq:dyndip_MF}) with $Q = 55\,\bar{\omega}_F$ are
reported in Fig.\ \ref{fig:dipo_comp} and compared with the numerical
results for the dependence of the mode frequencies and damping rates
on the anisotropy parameter. At $\lambda=0.05$ the gas is already very
close to the hydrodynamic regime in its motions along $\hat{z}$, since
the value of the oscillation frequency obtained in the simulation is
close to $\omega_{\text{hd},i}$ in spite of the low concentration of
impurities.  The overall agreement between the simulation and the
model is fairly good, and the deviations may be attributed to
low-temperature effects and to shape deformations of the distributions
not entering Eqs.\ (\ref{eq:dyndip_MF}).

Finally, it is worthwhile stressing that even though the effects on
the dipolar oscillations due to the scattering with impurities are
small, they can be greatly increased by tuning the experimental
parameters. For instance, if we consider the parameters of the recent
experiment by Takasu {\em et al.}  \cite{Takasu2003} and take 100
$^{174}$Yb atoms as impurities, the difference between the oscillation
frequencies along the axial and radial directions as shown in Fig.\
\ref{fig:dipo_comp} will be of about $75\%$ in the limit of small
$\lambda$. Therefore, we may expect that the effects here described
will become easily observable in the near future when mixtures of
$^{40}$K and $^{174}$Yb will be experimentally realized.

\subsection{Monopolar and quadrupolar oscillations}
Dipolar modes are a simple example of collective modes
that can be experimentally analyzed and whose features expose the
collisional state of the gas. As illustrated by the classical model,
these modes can be viewed mainly as coupled oscillations of the
centers of mass of the two clouds in the absence of appreciable
deformations of their shapes. Other collective modes can be explored
by deforming the clouds in different ways. The lowest-frequency modes
of this type are the surface monopolar ($l=0$) and quadrupolar ($l=2$)
modes, and these can be excited by small deformations of suitable
symmetry. While in an isotropic trap the monopolar mode and the
$\ell_z=0$ quadrupolar mode can be independently excited, in an
axially symmetric trap the angular momentum is not a good quantum
number and the two oscillations with $\ell_z=0$ are coupled to each
other.

\begin{figure}
\includegraphics[width=\linewidth]{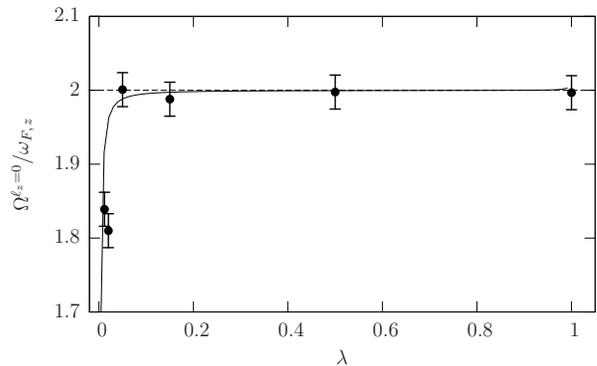}
\caption{\label{fig:quadrupole}Frequency $\Omega_0$ (in units of
$\omega_{F,\,z}$) of the lower oscillation mode with $\ell_z=0$, as a
function of $\lambda$ (symbols with error bars giving standard
deviations). The solid line shows the lower $\ell_z=0$ frequency for a
interacting fermion system with $\bar{\omega}_F\,\tau_{mq}=15$ (see
text).}
\end{figure}

As the collisionality of the fermion gas is increased, the transition
to the hydrodynamic regime in strongly elongated traps manifests
itself through changes in the $\ell_z=0$ mode frequencies from
$2\omega_{F,z}$ to $\Omega_{\text{hd}} = \sqrt{12/5}\,\omega_{F,z}$
and from $2\omega_{F, \perp}$ to $\sqrt{10/3}\,\omega_{F,\perp}$
\cite{Vichi2000a, Griffin1997a, Guery-Odelin1999b}. A similar
transition can be expected in the present system, where the impurities
act to transfer momentum between noninteracting fermions.  We have
monitored several dynamical averages of the fermion cloud in the
course of the simulation and in particular we have analyzed the time
evolution of $\chi(t)=\langle 2v_{F,z}^2-v_{F,\perp}^2\rangle$ after
compression of the cloud density by about 5\%. This average measures
the anisotropy of the velocity distribution and is the responsible for
the coupling between monopole and quadrupole excitations
\cite{Guery-Odelin1999b}. At variance from the dipole modes, in the
monopole and quadrupole modes the radial and axial motions are
strongly coupled and this requires that we simultaneously follow the
dynamics of the gas on two quite different time scales. In Fig.\
\ref{fig:quadrupole} we plot the value of the lower $\ell_z=0$ mode
frequency obtained from the Fourier transform of $\chi(t)$. At very
large anisotropy this peak frequency suddenly drops from the
collisionless value $2\omega_{F,z}$ towards the hydrodynamic value
$\sqrt{12/5}\,\omega_{F,z}$, whereas the higher peak frequency (not
shown in Fig. \ref{fig:quadrupole}) stops at its collisionless value
$2\omega_{F,\perp}$. Again collisionless and hydrodynamic behaviors
coexist in a strongly elongated trap.

 The behavior of monopolar and quadrupolar modes with finite
relaxation time $\tau_{mq}$ in a gas of interacting fermions inside an
anisotropic trap has been studied by Vichi \cite{Vichi2000a}, who
derived an implicit polynomial equation predicting a very steep
downturn of the lower mode frequency in systems with
$\bar{\omega}_F\,\tau_{mq} \gg 1$. The solid line in Fig.\
\ref{fig:quadrupole} shows the result of fitting Vichi's model to our data
with the choice $\bar{\omega}_F\,\tau_{mq}=15$.  The sharp transition
to the collisional regime is due to the fact that the impurities
mediate the fermion-fermion scattering and lead to an effective
relaxation time which is larger than that involved in the
impurity-fermion scattering.  Even though the effect is limited to
large anisotropies, it could become observable at moderate
anisotropies by increasing the numbers of particles or the strength of
the boson-fermion repulsion.

\section{\label{sec:expan}Expansion dynamics}

Many experiments extract information on the properties of an 
ultracold trapped gas after it been allowed to undergo free expansion. 
This improves the spatial resolution of {\em in-situ}
measurements and also gives access to the momentum distribution of the atoms. 
A fermion gas in the collisionless regime becomes spatially
isotropic as it expands, regardless of the initial anisotropy of its
density profile, whereas in the hydrodynamic regime the density
profile inverts its aspect ratio during expansion.
Furthermore, since the number of collisions diminishes as the gas
expands, the hydrodynamic picture will become invalid when there are not
enough collisions to sustain local equilibrium.

\begin{figure}
\includegraphics[width=\columnwidth,clip=true]{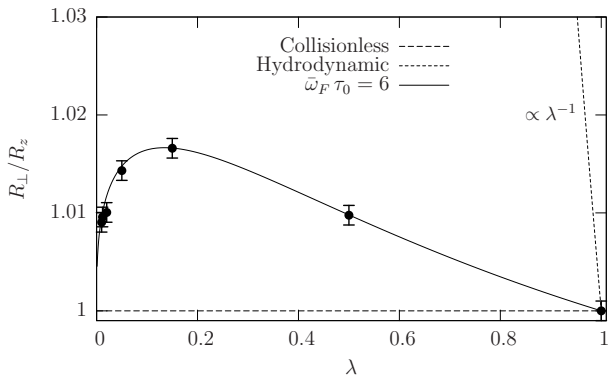}
\caption{\label{fig_expa}Asymptotic aspect ratio $R_{\perp}/R_z$ 
of the expanded
fermion cloud as a function of the trap anisotropy $\lambda$.  The
error bars are estimated from standard deviations in the fitting process. 
The solid line is the solution of Eq.\
(\ref{eq:stringari}) with $\tau_0=6/\bar{\omega}_F$.
The collisionless and hydrodynamic limits are shown by the dashed and
short-dashed lines, respectively.}
\end{figure}

In the numerical simulation of the expansion, starting from the
equilibrium density profiles of the trapped gaseous mixture we switch off the
confinement and allow evolution according to the
VLE. During the evolution we adaptively change the size of the
computational domain in order to ensure that all particles are
inside it. This permits us to study the free expansion of the 
cloud for long periods of time up to expansion ratios $b_i(t) =
R_i(t)/R_i(0)$ of a few hundreds, with $R_i(t)=\langle
x_{F,i}^2(t)\rangle^{1/2}$ being the width of the cloud in the
$i$-direction.  In Fig. \ref{fig_expa} we show the asymptotic value 
$R_{\perp}/R_z$ of the aspect ratio of the fermion cloud as a function of
the anisotropy $\lambda$.  
The aspect ratio in the case of isotropic or strongly
anisotropic confinement is equal to unity as in the
collisionless regime, while for intermediate values of $\lambda$ the
scattering against the impurities makes $R_{\perp}/R_{z}$ deviate from
unity.  This behavior can be seen as a consequence of the competition
between the time scales associated to axial and radial motions, as
explained below.

\begin{figure}
\includegraphics[width=0.9\columnwidth,clip=true]{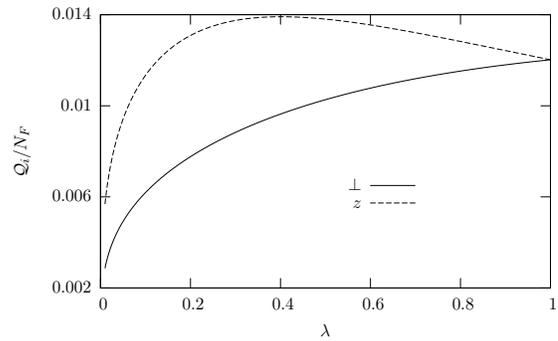}
\caption{\label{conticino_coll} Numbers
${\cal Q}_\perp/N_F$ (solid line) and ${\cal Q}_z/N_F$ (dashed line) 
of collisions per fermion, as
functions of the anisotropy parameter $\lambda$.}
\end{figure}

We have evaluated the total numbers ${\cal Q}_\perp$ and
${\cal Q}_z$ of collisions occurring during a lapse of time from $t=0$ to
$t_\perp=1/\omega_{F,\perp}$ and to $t_z = 1/\omega_{F,z}$,
respectively.  Assuming that the collision rate $Q(t)$ scales in time
with the volume occupied by the gas and that the expansion dynamics is
close to that of a collisionless system, $\mathcal{Q}_{\perp}$
and $\mathcal{Q}_{z}$ can be written as
\begin{eqnarray}
{\cal Q}_{\perp,z}&=&\int_0^{t_\perp,t_z}dt\,Q(t)\\& =&
Q(0)\int_{0}^{t_\perp,t_z}dt\,
(1+\omega_{F,\perp}^2t^2)^{-1}(1+\omega_{F,z}^2t^2)^{-1/2} .\nonumber
\label{venerdi}
\end{eqnarray}
In Eq.\ (\ref{venerdi}) we have taken $Q(t)= Q(0)/(b_\perp^2 b_z)$ and
set $b_i(t)=(1+\omega_{F,i}^2\,t^2)^{1/2}$ \cite{Pedri2003a}. The
integral in Eq.\ (\ref{venerdi}) is straightforward and yields the
results shown in Fig.\ \ref{conticino_coll}.
The following points should be noted: (i) the
characteristic time $t_\perp$ for the radial expansion and 
consequently ${\cal Q}_\perp$ decrease
with increasing the anisotropy; and (ii) the integrand of Eq. (\ref{venerdi}) 
for ${\cal Q}_z$ becomes
negligible for $t_\perp\ll t\le t_z$ as the fermion density drops due
to rapid expansion in the radial direction, so that ${\cal Q}_z$ increases
at first with decreasing $\lambda$, reaches its maximum
at $\lambda\simeq 0.4$, and then rapidly drops
as the density of the expanding cloud rapidly vanishes.

The non-monotonic behavior of $R_{\perp}/R_z$ as a function of
$\lambda$ in Fig. \ref{fig_expa} can be understood from the features 
of $\mathcal{Q}_{\perp}$ and $\mathcal{Q}_z$ in Fig. \ref{conticino_coll}. 
The number $\mathcal{Q}_{\perp}$ of collisions
diminishes with decreasing $\lambda$ while $\mathcal{Q}_z$ increases,
so that collisions play different roles in the axial and radial expansion.
Both $\mathcal{Q}_{\perp}$ and
$\mathcal{Q}_z$ are dropping at large anisotropies, so that
collisionless behavior is emerging in the ballistic expansion from a
 strongly elongated trap.

The behavior of the expanding gas as a function of $\lambda$ can also
be analyzed by means of scaling equations. In this approach
\cite{Pedri2003a} the expansion of a fermion gas with finite
collisionality can be described by the equations
\begin{equation}
\left\{
\begin{aligned}
\ddot{b}_i - \omega_{F,i}^2\,\frac{\theta_i}{b_i} &= 0  \\
\dot{\theta}_i + 2 \frac{\dot{b}_i}{b_i}\,\theta_i^2 &=
\frac{1}{\tau(\theta_i,b_i)}\left(\theta_i -
\frac{1}{3}\sum_j\theta_j\right)\,
\end{aligned}
\right.
\label{eq:stringari}
\end{equation}
where $\theta_i(t)$ are effective temperatures along the two
spatial directions, $\tau(\theta_i,b_i) = \tau_0\,(\prod_j b_j)
(\frac{1}{3}\sum_k\theta_k)^{-1/2}$ and $\tau_0$ is the collision time
at $t=0$. The collisionless and hydrodynamic limits correspond to
taking $\tau=\infty$ and $\tau=0$ in Eqs.\ (\ref{eq:stringari}),
respectively.

We have taken the collision time $\tau_0$ in Eqs.\
(\ref{eq:stringari}) as a fitting parameter to represent the indirect
fermion-fermion scattering induced by the impurities.  The solution of
Eqs. (\ref{eq:stringari}) with the choice $\tau_0=6/\bar{\omega}_F$
reproduces the non-monotonic behavior of the aspect ratio as shown in
Fig.\ \ref{fig_expa}.  Finally, we have checked that the qualitative
behavior shown in Fig. \ref{conticino_coll} for ${\cal Q}_{\perp}$ and
${\cal Q}_z$ does not depend on the assumption of a collisionless
evolution of the gas and persists on describing its expansion with any
value of $\tau_0$.  The magnitude of the deviation from unity of the
aspect ratio of the expanding fermions depends on the value of
$\tau_0$ and can be increased by changing the fermion-impurity
interaction {\em via} Feshbach resonances \cite{Inouye2004a} or by
performing an out-of-equilibrium experiment as described in Ref.\
\cite{Capuzzi2004b}.

\section{\label{sec:summary}Summary and concluding remarks}

We have studied the low-lying surface modes and the ballistic
expansion of a spin-polarized Fermi gas interacting with thermal
impurities as functions of the anisotropy of its confinement in a
cigar-shaped harmonic trap.  We have solved for this purpose the
Vlasov-Landau equations for the dynamics of the mixture and compared
the results with simple scaling equations containing collision-time
fitting parameters.  The results show that for large anisotropies a
collisionless behavior in the radial dipolar oscillations and a
hydrodynamic behavior in the axial ones are simultaneously
established. For monopolar and quadrupolar excitations we have
observed a collisionless spectrum irrespectively of the strength of
the anisotropy parameter, except for extremely large anisotropies
where the frequency of the lower $\ell_z=0$ mode decreases towards the
hydrodynamic value.

On the other hand, during ballistic expansion the two different time
scales for radial and axial motions enter into competition. The result
is that the expansion of a strongly anisotropic cloud is essentially
collisionless due to the rapid drop of the particle density, whereas
at intermediate values of the anisotropy the aspect ratio is sensitive
to the collisions.

The analysis presented here has only concerned slightly doped fermions
in a collisionless-to-intermediate scattering regime as, e.g., that
attained in the experiments at LENS. The effects of the trap
anisotropy could be further enhanced exploiting the rich variety of
experimental set-ups and trapped isotopes available. It would be also
interesting to extend our work to strong-interaction situations among
fermions and impurities or among fermions in two-component Fermi gases
approaching the unitary limit.

\acknowledgments

This work has been partially supported by an Advanced Research
Initiative of Scuola Normale Superiore di Pisa and by the Istituto
Nazionale di Fisica della Materia within the Initiative ``Calcolo
Parallelo''.

\appendix


\begin{thebibliography}{27}
\expandafter\ifx\csname natexlab\endcsname\relax\def\natexlab#1{#1}\fi
\expandafter\ifx\csname bibnamefont\endcsname\relax
  \def\bibnamefont#1{#1}\fi
\expandafter\ifx\csname bibfnamefont\endcsname\relax
  \def\bibfnamefont#1{#1}\fi
\expandafter\ifx\csname citenamefont\endcsname\relax
  \def\citenamefont#1{#1}\fi
\expandafter\ifx\csname url\endcsname\relax
  \def\url#1{\texttt{#1}}\fi
\expandafter\ifx\csname urlprefix\endcsname\relax\def\urlprefix{URL }\fi
\providecommand{\bibinfo}[2]{#2}
\providecommand{\eprint}[2][]{\url{#2}}

\bibitem[{\citenamefont{Minguzzi et~al.}(2004)\citenamefont{Minguzzi, Succi,
  Toschi, Tosi, and Vignolo}}]{Minguzzi2004a}
\bibinfo{author}{\bibfnamefont{A.}~\bibnamefont{Minguzzi}},
  \bibinfo{author}{\bibfnamefont{S.}~\bibnamefont{Succi}},
  \bibinfo{author}{\bibfnamefont{F.}~\bibnamefont{Toschi}},
  \bibinfo{author}{\bibfnamefont{M.~P.} \bibnamefont{Tosi}}, \bibnamefont{and}
  \bibinfo{author}{\bibfnamefont{P.}~\bibnamefont{Vignolo}},
  \bibinfo{journal}{Phys. Rep.} \textbf{\bibinfo{volume}{395}},
  \bibinfo{pages}{223} (\bibinfo{year}{2004}).

\bibitem[{\citenamefont{Anderson et~al.}(1995)\citenamefont{Anderson, Ensher,
  Matthews, Wieman, and Cornell}}]{Anderson1995a}
\bibinfo{author}{\bibfnamefont{M.~H.} \bibnamefont{Anderson}},
  \bibinfo{author}{\bibfnamefont{J.~R.} \bibnamefont{Ensher}},
  \bibinfo{author}{\bibfnamefont{M.~R.} \bibnamefont{Matthews}},
  \bibinfo{author}{\bibfnamefont{C.~E.} \bibnamefont{Wieman}},
  \bibnamefont{and} \bibinfo{author}{\bibfnamefont{E.~A.}
  \bibnamefont{Cornell}}, \bibinfo{journal}{Science}
  \textbf{\bibinfo{volume}{269}}, \bibinfo{pages}{198} (\bibinfo{year}{1995}).

\bibitem[{\citenamefont{Bradley et~al.}(1995)\citenamefont{Bradley, Sackett,
  Tollett, and Hulet}}]{Bradley1995a}
\bibinfo{author}{\bibfnamefont{C.~C.} \bibnamefont{Bradley}},
  \bibinfo{author}{\bibfnamefont{C.~A.} \bibnamefont{Sackett}},
  \bibinfo{author}{\bibfnamefont{J.~J.} \bibnamefont{Tollett}},
  \bibnamefont{and} \bibinfo{author}{\bibfnamefont{R.~G.} \bibnamefont{Hulet}},
  \bibinfo{journal}{Phys. Rev. Lett.} \textbf{\bibinfo{volume}{75}},
  \bibinfo{pages}{1687} (\bibinfo{year}{1995}).

\bibitem[{\citenamefont{Davis et~al.}(1995)\citenamefont{Davis, Mewes, Andrews,
  van Druten, Durfee, Kurn, and Ketterle}}]{Davis1995b}
\bibinfo{author}{\bibfnamefont{K.~B.} \bibnamefont{Davis}},
  \bibinfo{author}{\bibfnamefont{M.~O.} \bibnamefont{Mewes}},
  \bibinfo{author}{\bibfnamefont{M.~R.} \bibnamefont{Andrews}},
  \bibinfo{author}{\bibfnamefont{N.~J.} \bibnamefont{van Druten}},
  \bibinfo{author}{\bibfnamefont{D.~S.} \bibnamefont{Durfee}},
  \bibinfo{author}{\bibfnamefont{D.~M.} \bibnamefont{Kurn}}, \bibnamefont{and}
  \bibinfo{author}{\bibfnamefont{W.}~\bibnamefont{Ketterle}},
  \bibinfo{journal}{Phys. Rev. Lett.} \textbf{\bibinfo{volume}{75}},
  \bibinfo{pages}{3969} (\bibinfo{year}{1995}).

\bibitem[{\citenamefont{DeMarco and Jin}(1999)}]{DeMarco1999a}
\bibinfo{author}{\bibfnamefont{B.}~\bibnamefont{DeMarco}} \bibnamefont{and}
  \bibinfo{author}{\bibfnamefont{D.~S.} \bibnamefont{Jin}},
  \bibinfo{journal}{Science} \textbf{\bibinfo{volume}{285}},
  \bibinfo{pages}{1703} (\bibinfo{year}{1999}).

\bibitem[{\citenamefont{Truscott et~al.}(2001)\citenamefont{Truscott, Strecker,
  McAlexander, Partridge, and Hulet}}]{Truscott2001a}
\bibinfo{author}{\bibfnamefont{A.~G.} \bibnamefont{Truscott}},
  \bibinfo{author}{\bibfnamefont{K.~E.} \bibnamefont{Strecker}},
  \bibinfo{author}{\bibfnamefont{W.~I.} \bibnamefont{McAlexander}},
  \bibinfo{author}{\bibfnamefont{G.~B.} \bibnamefont{Partridge}},
  \bibnamefont{and} \bibinfo{author}{\bibfnamefont{R.~G.} \bibnamefont{Hulet}},
  \bibinfo{journal}{Science} \textbf{\bibinfo{volume}{291}},
  \bibinfo{pages}{2570} (\bibinfo{year}{2001}).

\bibitem[{\citenamefont{Schreck et~al.}(2001)\citenamefont{Schreck, Khaykovich,
  Corwin, Ferrari, Bourdel, Cubizolles, and Salomon}}]{Schreck2001a}
\bibinfo{author}{\bibfnamefont{F.}~\bibnamefont{Schreck}},
  \bibinfo{author}{\bibfnamefont{L.}~\bibnamefont{Khaykovich}},
  \bibinfo{author}{\bibfnamefont{K.~L.} \bibnamefont{Corwin}},
  \bibinfo{author}{\bibfnamefont{G.}~\bibnamefont{Ferrari}},
  \bibinfo{author}{\bibfnamefont{T.}~\bibnamefont{Bourdel}},
  \bibinfo{author}{\bibfnamefont{J.}~\bibnamefont{Cubizolles}},
  \bibnamefont{and} \bibinfo{author}{\bibfnamefont{C.}~\bibnamefont{Salomon}},
  \bibinfo{journal}{Phys. Rev. Lett.} \textbf{\bibinfo{volume}{87}},
  \bibinfo{pages}{080403} (\bibinfo{year}{2001}).

\bibitem[{\citenamefont{Goldwin et~al.}(2002)\citenamefont{Goldwin, Papp,
  DeMarco, and Jin}}]{Goldwin2002a}
\bibinfo{author}{\bibfnamefont{J.}~\bibnamefont{Goldwin}},
  \bibinfo{author}{\bibfnamefont{S.~B.} \bibnamefont{Papp}},
  \bibinfo{author}{\bibfnamefont{B.}~\bibnamefont{DeMarco}}, \bibnamefont{and}
  \bibinfo{author}{\bibfnamefont{D.~S.} \bibnamefont{Jin}},
  \bibinfo{journal}{Phys. Rev. A} \textbf{\bibinfo{volume}{65}},
  \bibinfo{pages}{021402} (\bibinfo{year}{2002}).

\bibitem[{\citenamefont{O'Hara et~al.}(2002)\citenamefont{O'Hara, Hemmer, Gehm,
  Granade, and Thomas}}]{Ohara2002b}
\bibinfo{author}{\bibfnamefont{K.~M.} \bibnamefont{O'Hara}},
  \bibinfo{author}{\bibfnamefont{S.~L.} \bibnamefont{Hemmer}},
  \bibinfo{author}{\bibfnamefont{M.~E.} \bibnamefont{Gehm}},
  \bibinfo{author}{\bibfnamefont{S.~R.} \bibnamefont{Granade}},
  \bibnamefont{and} \bibinfo{author}{\bibfnamefont{J.~E.}
  \bibnamefont{Thomas}}, \bibinfo{journal}{Science}
  \textbf{\bibinfo{volume}{298}}, \bibinfo{pages}{2179} (\bibinfo{year}{2002}).

\bibitem[{\citenamefont{Greiner et~al.}(2003)\citenamefont{Greiner, Regal, and
  Jin}}]{Greiner2003a}
\bibinfo{author}{\bibfnamefont{M.}~\bibnamefont{Greiner}},
  \bibinfo{author}{\bibfnamefont{C.~A.} \bibnamefont{Regal}}, \bibnamefont{and}
  \bibinfo{author}{\bibfnamefont{D.~S.} \bibnamefont{Jin}},
  \bibinfo{journal}{Nature} \textbf{\bibinfo{volume}{426}},
  \bibinfo{pages}{537} (\bibinfo{year}{2003}).

\bibitem[{\citenamefont{Chin et~al.}(2004)\citenamefont{Chin, Bartenstein,
  Altmeyer, Riedl, Jochim, {Hecker Denschlag}, and Grimm}}]{Chin2004a}
\bibinfo{author}{\bibfnamefont{C.}~\bibnamefont{Chin}},
  \bibinfo{author}{\bibfnamefont{M.}~\bibnamefont{Bartenstein}},
  \bibinfo{author}{\bibfnamefont{A.}~\bibnamefont{Altmeyer}},
  \bibinfo{author}{\bibfnamefont{S.}~\bibnamefont{Riedl}},
  \bibinfo{author}{\bibfnamefont{S.}~\bibnamefont{Jochim}},
  \bibinfo{author}{\bibfnamefont{J.}~\bibnamefont{{Hecker Denschlag}}},
  \bibnamefont{and} \bibinfo{author}{\bibfnamefont{R.}~\bibnamefont{Grimm}},
  \bibinfo{journal}{Science} \textbf{\bibinfo{volume}{305}},
  \bibinfo{pages}{1128} (\bibinfo{year}{2004}).

\bibitem[{\citenamefont{Graham and Walls}(1998)}]{Graham1998a}
\bibinfo{author}{\bibfnamefont{R.}~\bibnamefont{Graham}} \bibnamefont{and}
  \bibinfo{author}{\bibfnamefont{D.}~\bibnamefont{Walls}},
  \bibinfo{journal}{Phys. Rev. A} \textbf{\bibinfo{volume}{57}},
  \bibinfo{pages}{484} (\bibinfo{year}{1998}).

\bibitem[{\citenamefont{Capuzzi
  et~al.}(2003{\natexlab{a}})\citenamefont{Capuzzi, Minguzzi, and
  Tosi}}]{capuzzi2003a}
\bibinfo{author}{\bibfnamefont{P.}~\bibnamefont{Capuzzi}},
  \bibinfo{author}{\bibfnamefont{A.}~\bibnamefont{Minguzzi}}, \bibnamefont{and}
  \bibinfo{author}{\bibfnamefont{M.~P.} \bibnamefont{Tosi}},
  \bibinfo{journal}{Phys. Rev. A} \textbf{\bibinfo{volume}{67}},
  \bibinfo{pages}{053605} (\bibinfo{year}{2003}{\natexlab{a}}).

\bibitem[{\citenamefont{Capuzzi
  et~al.}(2003{\natexlab{b}})\citenamefont{Capuzzi, Minguzzi, and
  Tosi}}]{capuzzi2003b}
\bibinfo{author}{\bibfnamefont{P.}~\bibnamefont{Capuzzi}},
  \bibinfo{author}{\bibfnamefont{A.}~\bibnamefont{Minguzzi}}, \bibnamefont{and}
  \bibinfo{author}{\bibfnamefont{M.~P.} \bibnamefont{Tosi}},
  \bibinfo{journal}{Phys. Rev. A} \textbf{\bibinfo{volume}{68}},
  \bibinfo{pages}{033605} (\bibinfo{year}{2003}{\natexlab{b}}).

\bibitem[{\citenamefont{Kinast et~al.}(2004{\natexlab{a}})\citenamefont{Kinast,
  Hemmer, Gehm, Turlapov, and Thomas}}]{Kinast2004a}
\bibinfo{author}{\bibfnamefont{J.}~\bibnamefont{Kinast}},
  \bibinfo{author}{\bibfnamefont{S.~L.} \bibnamefont{Hemmer}},
  \bibinfo{author}{\bibfnamefont{M.~E.} \bibnamefont{Gehm}},
  \bibinfo{author}{\bibfnamefont{A.}~\bibnamefont{Turlapov}}, \bibnamefont{and}
  \bibinfo{author}{\bibfnamefont{J.~E.} \bibnamefont{Thomas}},
  \bibinfo{journal}{Phys. Rev. Lett.} \textbf{\bibinfo{volume}{92}},
  \bibinfo{pages}{150402} (\bibinfo{year}{2004}{\natexlab{a}}).

\bibitem[{\citenamefont{Bartenstein et~al.}(2004)\citenamefont{Bartenstein,
  Altmeyer, Riedl, Jochim, Chin, Hecker~Denschlag, and
  Grimm}}]{Bartenstein2004a}
\bibinfo{author}{\bibfnamefont{M.}~\bibnamefont{Bartenstein}},
  \bibinfo{author}{\bibfnamefont{A.}~\bibnamefont{Altmeyer}},
  \bibinfo{author}{\bibfnamefont{S.}~\bibnamefont{Riedl}},
  \bibinfo{author}{\bibfnamefont{S.}~\bibnamefont{Jochim}},
  \bibinfo{author}{\bibfnamefont{C.}~\bibnamefont{Chin}},
  \bibinfo{author}{\bibfnamefont{J.}~\bibnamefont{Hecker~Denschlag}},
  \bibnamefont{and} \bibinfo{author}{\bibfnamefont{R.}~\bibnamefont{Grimm}},
  \bibinfo{journal}{Phys. Rev. Lett.} \textbf{\bibinfo{volume}{92}},
  \bibinfo{pages}{203201} (\bibinfo{year}{2004}).

\bibitem[{\citenamefont{Kinast et~al.}(2004{\natexlab{b}})\citenamefont{Kinast,
  Turlapov, and Thomas}}]{Kinast2004b}
\bibinfo{author}{\bibfnamefont{J.}~\bibnamefont{Kinast}},
  \bibinfo{author}{\bibfnamefont{A.}~\bibnamefont{Turlapov}}, \bibnamefont{and}
  \bibinfo{author}{\bibfnamefont{J.~E.} \bibnamefont{Thomas}}
  (\bibinfo{year}{2004}{\natexlab{b}}), \eprint{cond-mat/0408634}.

\bibitem[{\citenamefont{Gensemer and Jin}(2001)}]{Gensemer2001a}
\bibinfo{author}{\bibfnamefont{S.~D.} \bibnamefont{Gensemer}} \bibnamefont{and}
  \bibinfo{author}{\bibfnamefont{D.~S.} \bibnamefont{Jin}},
  \bibinfo{journal}{Phys. Rev. Lett.} \textbf{\bibinfo{volume}{87}},
  \bibinfo{pages}{173201} (\bibinfo{year}{2001}).

\bibitem[{\citenamefont{Toschi et~al.}(2003)\citenamefont{Toschi, Vignolo,
  Succi, and Tosi}}]{Toschi2003a}
\bibinfo{author}{\bibfnamefont{F.}~\bibnamefont{Toschi}},
  \bibinfo{author}{\bibfnamefont{P.}~\bibnamefont{Vignolo}},
  \bibinfo{author}{\bibfnamefont{S.}~\bibnamefont{Succi}}, \bibnamefont{and}
  \bibinfo{author}{\bibfnamefont{M.~P.} \bibnamefont{Tosi}},
  \bibinfo{journal}{Phys. Rev. A} \textbf{\bibinfo{volume}{67}},
  \bibinfo{pages}{041605} (\bibinfo{year}{2003}).

\bibitem[{\citenamefont{Capuzzi et~al.}(2004)\citenamefont{Capuzzi, Vignolo,
  Toschi, Succi, and Tosi}}]{Capuzzi2004b}
\bibinfo{author}{\bibfnamefont{P.}~\bibnamefont{Capuzzi}},
  \bibinfo{author}{\bibfnamefont{P.}~\bibnamefont{Vignolo}},
  \bibinfo{author}{\bibfnamefont{F.}~\bibnamefont{Toschi}},
  \bibinfo{author}{\bibfnamefont{S.}~\bibnamefont{Succi}}, \bibnamefont{and}
  \bibinfo{author}{\bibfnamefont{M.~P.} \bibnamefont{Tosi}},
  \bibinfo{journal}{Phys. Rev. A} \textbf{\bibinfo{volume}{70}},
  \bibinfo{pages}{043623} (\bibinfo{year}{2004}).

\bibitem[{\citenamefont{Succi et~al.}(2004)\citenamefont{Succi, Toschi,
  Capuzzi, Vignolo, and Tosi}}]{Succi2003a}
\bibinfo{author}{\bibfnamefont{S.}~\bibnamefont{Succi}},
  \bibinfo{author}{\bibfnamefont{F.}~\bibnamefont{Toschi}},
  \bibinfo{author}{\bibfnamefont{P.}~\bibnamefont{Capuzzi}},
  \bibinfo{author}{\bibfnamefont{P.}~\bibnamefont{Vignolo}}, \bibnamefont{and}
  \bibinfo{author}{\bibfnamefont{M.~P.} \bibnamefont{Tosi}},
  \bibinfo{journal}{Phil. Trans. R. Soc. London A}
  \textbf{\bibinfo{volume}{362}}, \bibinfo{pages}{1605} (\bibinfo{year}{2004}).

\bibitem[{\citenamefont{Ferlaino et~al.}(2003)\citenamefont{Ferlaino, Brecha,
  Hannaford, Riboli, Roati, Modugno, and Inguscio}}]{Ferlaino2003a}
\bibinfo{author}{\bibfnamefont{F.}~\bibnamefont{Ferlaino}},
  \bibinfo{author}{\bibfnamefont{R.~J.} \bibnamefont{Brecha}},
  \bibinfo{author}{\bibfnamefont{P.}~\bibnamefont{Hannaford}},
  \bibinfo{author}{\bibfnamefont{F.}~\bibnamefont{Riboli}},
  \bibinfo{author}{\bibfnamefont{G.}~\bibnamefont{Roati}},
  \bibinfo{author}{\bibfnamefont{G.}~\bibnamefont{Modugno}}, \bibnamefont{and}
  \bibinfo{author}{\bibfnamefont{M.}~\bibnamefont{Inguscio}},
  \bibinfo{journal}{J. Opt. B} \textbf{\bibinfo{volume}{5}},
  \bibinfo{pages}{S3} (\bibinfo{year}{2003}).



\bibitem[{\citenamefont{Vichi and Stringari}(1999)}]{Vichi1999a}
\bibinfo{author}{\bibfnamefont{L.}~\bibnamefont{Vichi}} \bibnamefont{and}
  \bibinfo{author}{\bibfnamefont{S.}~\bibnamefont{Stringari}},
  \bibinfo{journal}{Phys. Rev. A} \textbf{\bibinfo{volume}{60}},
  \bibinfo{pages}{4734} (\bibinfo{year}{1999}).

\bibitem{Takasu2003}Y. Takasu, K. Maki, K. Komori, T. Takano,
  K. Honda, M. Kumakura, T. Yabuzaki, and Y. Takahashi,
  Phys. Rev. Lett. {\bf 91}, 040404 (2003).

\bibitem[{\citenamefont{Vichi}(2000)}]{Vichi2000a}
\bibinfo{author}{\bibfnamefont{L.}~\bibnamefont{Vichi}}, \bibinfo{journal}{J.
  Low. Temp. Phys.} \textbf{\bibinfo{volume}{121}}, \bibinfo{pages}{177}
  (\bibinfo{year}{2000}).

\bibitem[{\citenamefont{Griffin et~al.}(1997)\citenamefont{Griffin, Wu, and
  Stringari}}]{Griffin1997a}
\bibinfo{author}{\bibfnamefont{A.}~\bibnamefont{Griffin}},
  \bibinfo{author}{\bibfnamefont{W.-C.} \bibnamefont{Wu}}, \bibnamefont{and}
  \bibinfo{author}{\bibfnamefont{S.}~\bibnamefont{Stringari}},
  \bibinfo{journal}{Phys. Rev. Lett.} \textbf{\bibinfo{volume}{78}},
  \bibinfo{pages}{1838} (\bibinfo{year}{1997}).

\bibitem[{\citenamefont{Gu\'ery-Odelin
  et~al.}(1999)\citenamefont{Gu\'ery-Odelin, Zambelli, Dalibard, and
  Stringari}}]{Guery-Odelin1999b}
\bibinfo{author}{\bibfnamefont{D.}~\bibnamefont{Gu\'ery-Odelin}},
  \bibinfo{author}{\bibfnamefont{F.}~\bibnamefont{Zambelli}},
  \bibinfo{author}{\bibfnamefont{J.}~\bibnamefont{Dalibard}}, \bibnamefont{and}
  \bibinfo{author}{\bibfnamefont{S.}~\bibnamefont{Stringari}},
  \bibinfo{journal}{Phys. Rev. A} \textbf{\bibinfo{volume}{60}},
  \bibinfo{pages}{4851} (\bibinfo{year}{1999}).

\bibitem[{\citenamefont{Pedri et~al.}(2003)\citenamefont{Pedri, Gu\'ery-Odelin,
  and Stringari}}]{Pedri2003a}
\bibinfo{author}{\bibfnamefont{P.}~\bibnamefont{Pedri}},
  \bibinfo{author}{\bibfnamefont{D.}~\bibnamefont{Gu\'ery-Odelin}},
  \bibnamefont{and}
  \bibinfo{author}{\bibfnamefont{S.}~\bibnamefont{Stringari}},
  \bibinfo{journal}{Phys. Rev. A} \textbf{\bibinfo{volume}{68}},
  \bibinfo{pages}{043608} (\bibinfo{year}{2003}).

\bibitem{Inouye2004a} S. Inouye, J. Goldwin, M. L. Olsen, J. L. Bohn,
  and D. S. Jin, Phys. Rev. Lett. {\bf 93}, 183201 (2004).
\end{thebibliography}

\end{document}